\documentclass{PoS}

\usepackage[]{natbib}
\usepackage{graphicx}
\usepackage{aas_macros}
\usepackage{multicol}

\title{Properties of the radio jet emission of
  four gamma-ray Narrow Line Seyfert 1 galaxies}

\ShortTitle{Properties of the radio jet emission of
  four gamma-ray NLSy1s }

\author{\speaker{E. Angelakis}, L. Fuhrmann, I. Myserlis, I. Nestoras, V. Karamanavis,
  T.~P.~Krichbaum, J. A. Zensus\\
         Max-Planck-Institut f\"ur Radioastronomie, Auf dem H\"ugel 69, Bonn 53121, DE\\
        E-mail: \email{eangelakis@mpifr.de}
}

\author{N. Marchili\\
        Dipartimento di Fisica ed Astronomia, Universita di Padova, via Marzolo 8, Padova 35131,
  IT\\
}

\author{L. Foschini\\
        Istituto Nazionale di Astrofisica, Osservatorio Astronomico di Brera, Via
        E. Bianchi, 46 23807, Merate, IT\\
}

\author{H. Ungerechts, A. Sievers\\
  Instituto de Radio Astronomia Milimetrica, Avenida Divina Pastora 7, Local 20, E 18012 Granada, ES \\
}

\abstract{The detection of $\gamma$ rays from a small number of Narrow Line Seyfert 1
  galaxies by the LAT instrument onboard {\sl Fermi} seriously challenged our
  understanding of AGN physics. Among the most important findings associated with their
  discovery has been the realisation that smaller-mass black holes seem to be hosted by
  these systems. Immediately after their discovery a radio multi-frequency monitoring
  campaign was initiated to understand their jet radio emission. Here the first results of
  the campaign are presented. The light curves and some first variability analyses are
  discussed, showing that the brightness temperatures and Doppler factors are
   moderate. The phenomenologies are typically blazar-like. The frequency domain on the
  other hand indicates intense spectral evolution and the variability patterns indicate
  mechanisms similar to those acting in the jets of BL Lacs and FSRQs. Finally, the linear
  polarisation also reveals the presence of a quiescent, optically thin jet in certain cases.}

\FullConference{Nuclei of Seyfert Galaxies and QSOs - Central Engine and Conditions of Star Formation,\\
		November 6-8, 2012\\
		Max-Planck-Insitut f\"ur Radioastronomie (MPIfR), Bonn,  			Germany}

\begin{document}

\section{Introduction}
The $\gamma$-ray emitting classes of Active Galactic Nuclei (hereafter AGN) that were
thought to be there until 2008 were {\sl blazars} -- collectively referring to {\sl BL
  Lac} objects (hereafter BL Lacs) and {\sl flat spectrum radio quasars} (hereafter FSRQs)
-- and {\sl radio galaxies}.
In 2009 the LAT instrument onboard the {\sl Fermi}-GST satellite detected high-energy
$\gamma$ rays ($E>100$\,MeV) from the Narrow Line Seyfert 1 galaxy (NLSy1) PMN\,J0948+0022
with a blazar-like SED \citep{2009ApJ...699..976A}.
The multi-wavelength campaigns that followed the discovery \citep{2009ApJ...707..727A,
  2009ApJ...699..976A, 2011AnA...528L..11G,2012A&A...548A.106F} and the study of
subsequently discovered NLSy1s \citep{2009ApJ...707L.142A} showed a clear blazar-like
behaviour, indicating the existence of a relativistic jet viewed at small angles.

These detections have serious impact on our understanding of AGN. They challenge the belief
that powerful relativistic jets form in giant elliptical galaxies since Seyferts mostly reside in disc galaxies
\citep{2007ApJ...658..815S}.  Furthermore, they have implications on the energy production
and dissipation at different scales because despite the fundamental differences between
radio loud NLSy1s and blazars -- in terms of black hole masses
($10^{6}-10^{8}$\,M$_{\odot}$ ) and accretion disc luminosities
($0.2-0.9$\,$L_\mathrm{Edd}$), their jets seem to be behaving similarly and share the same
properties. Hence they provide an excellent probe for studying the physics scaling towards
lower black hole masses (see review by \citealt{2006AJ....132..531K}).

Here first results of a ``single-dish'' multi-frequency monitoring are reviewed and
compared with those seen in typical AGNs. Focus is being put on the variability brightness
temperatures as estimates of the energetics of flaring events and the variability pattern
followed by the broad-band radio spectra as indicator of the variability and emission
mechanism at play. The data discussed here are already partly published
\citep{2012A&A...548A.106F} and partly in preparation for
publication. Early results of this study are discussed by \cite{2011nlsg.confE..26F}.


\section{Light curves and variability}
\label{sec:lcs}
\begin{figure}[] 
\centering
\begin{tabular}{cc}
\includegraphics[width=5cm,angle=-90]{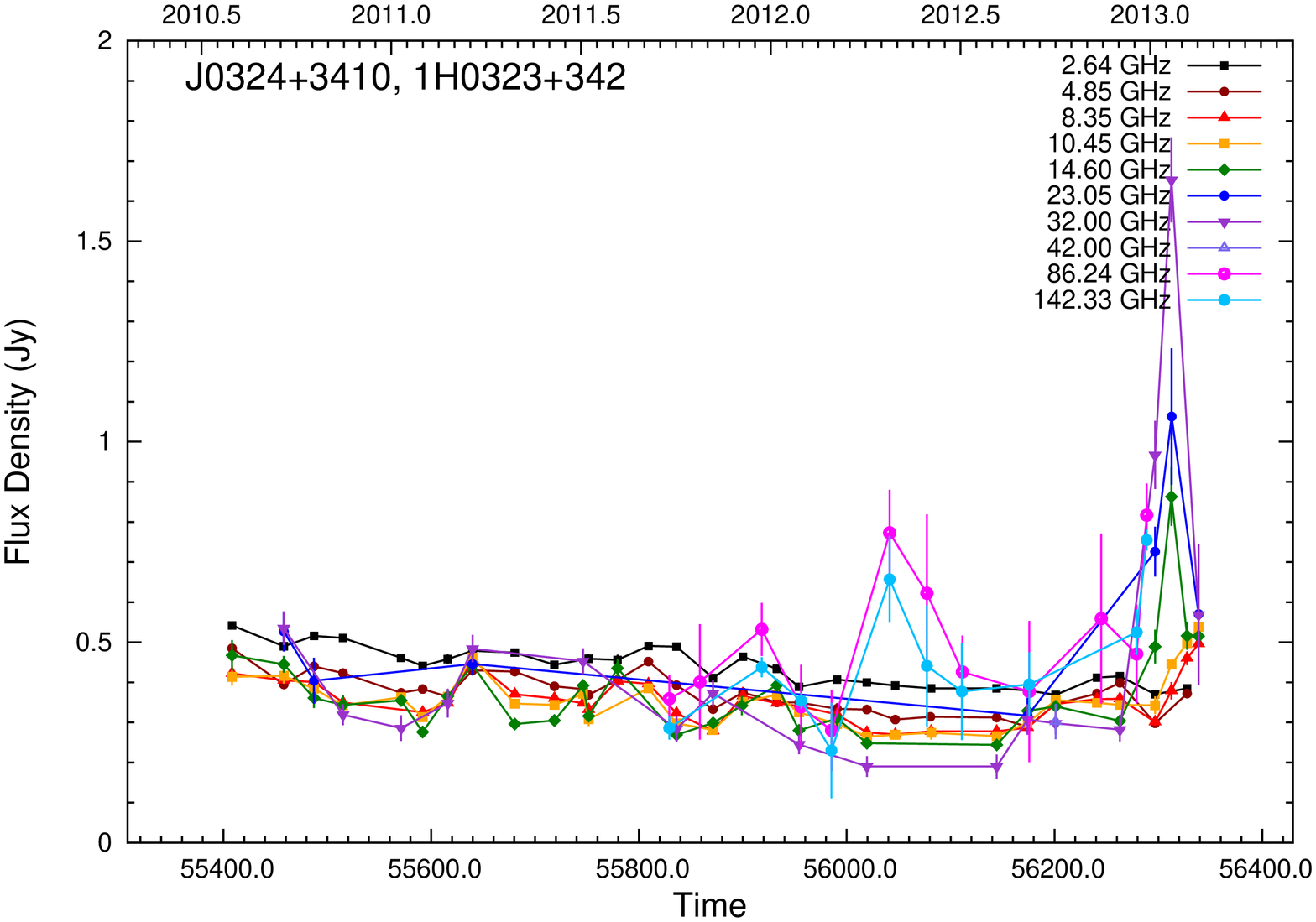}& 
\includegraphics[width=5cm,angle=-90]{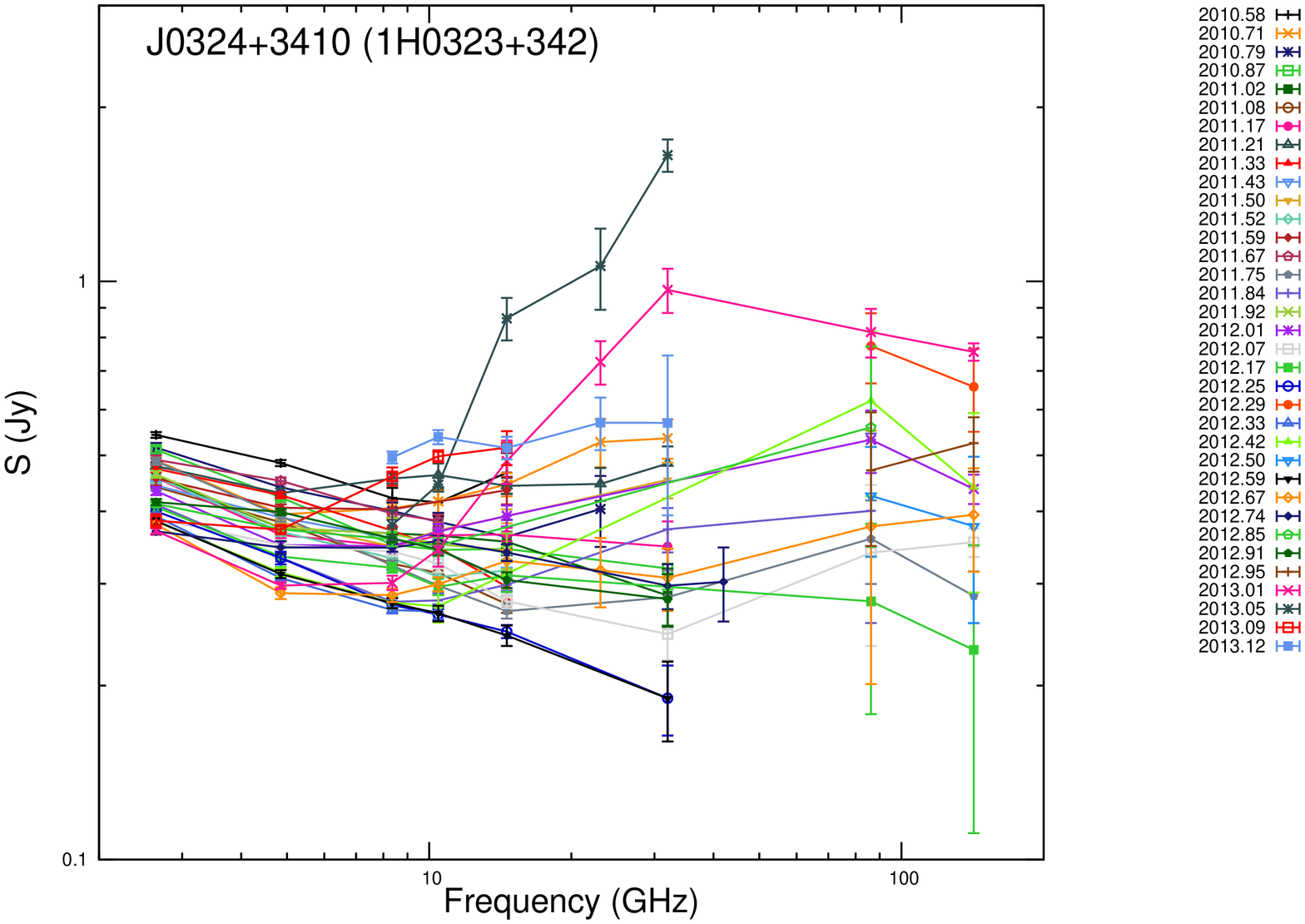}\\ 
\includegraphics[width=5cm,angle=-90]{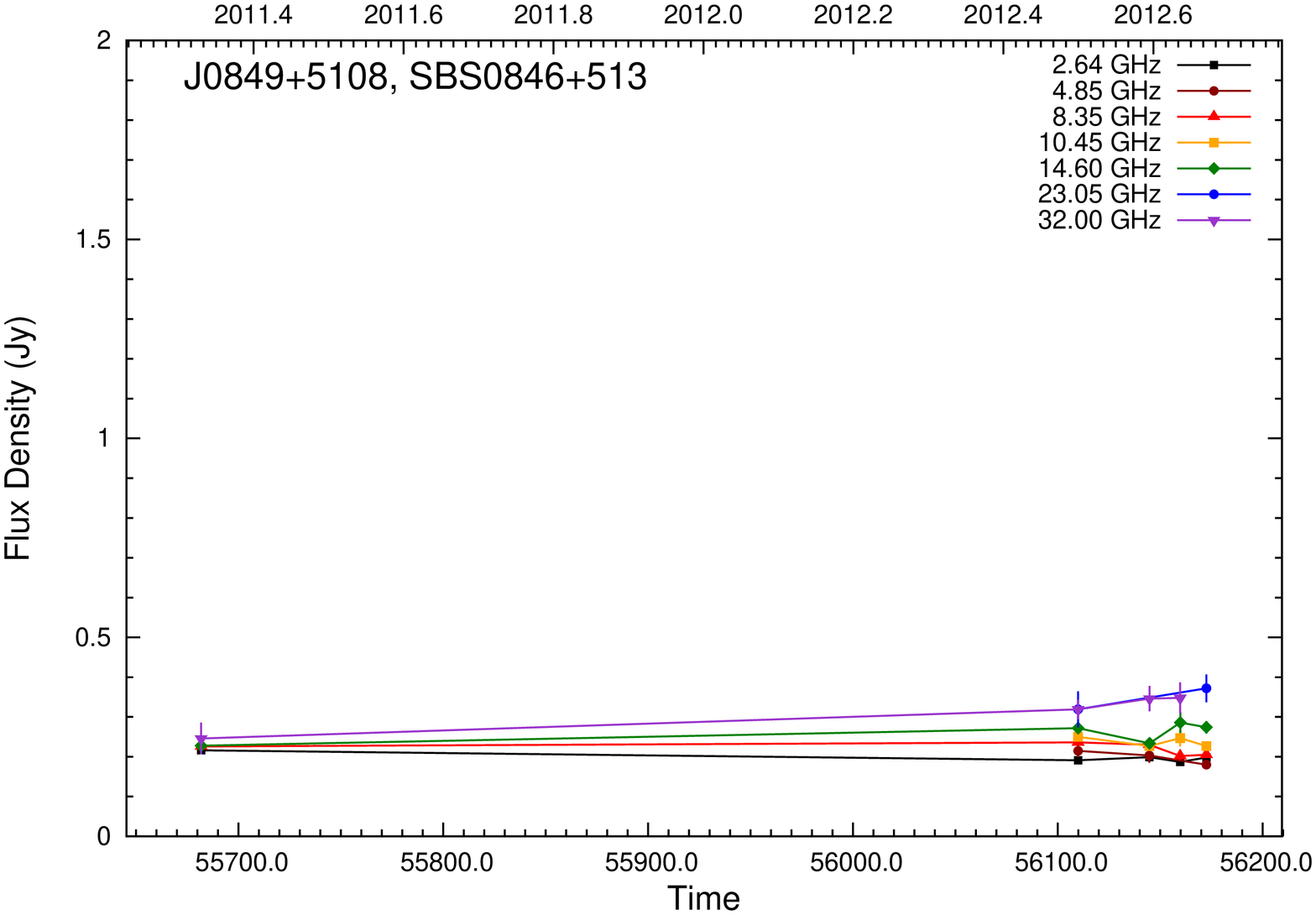}&
\includegraphics[width=5cm,angle=-90]{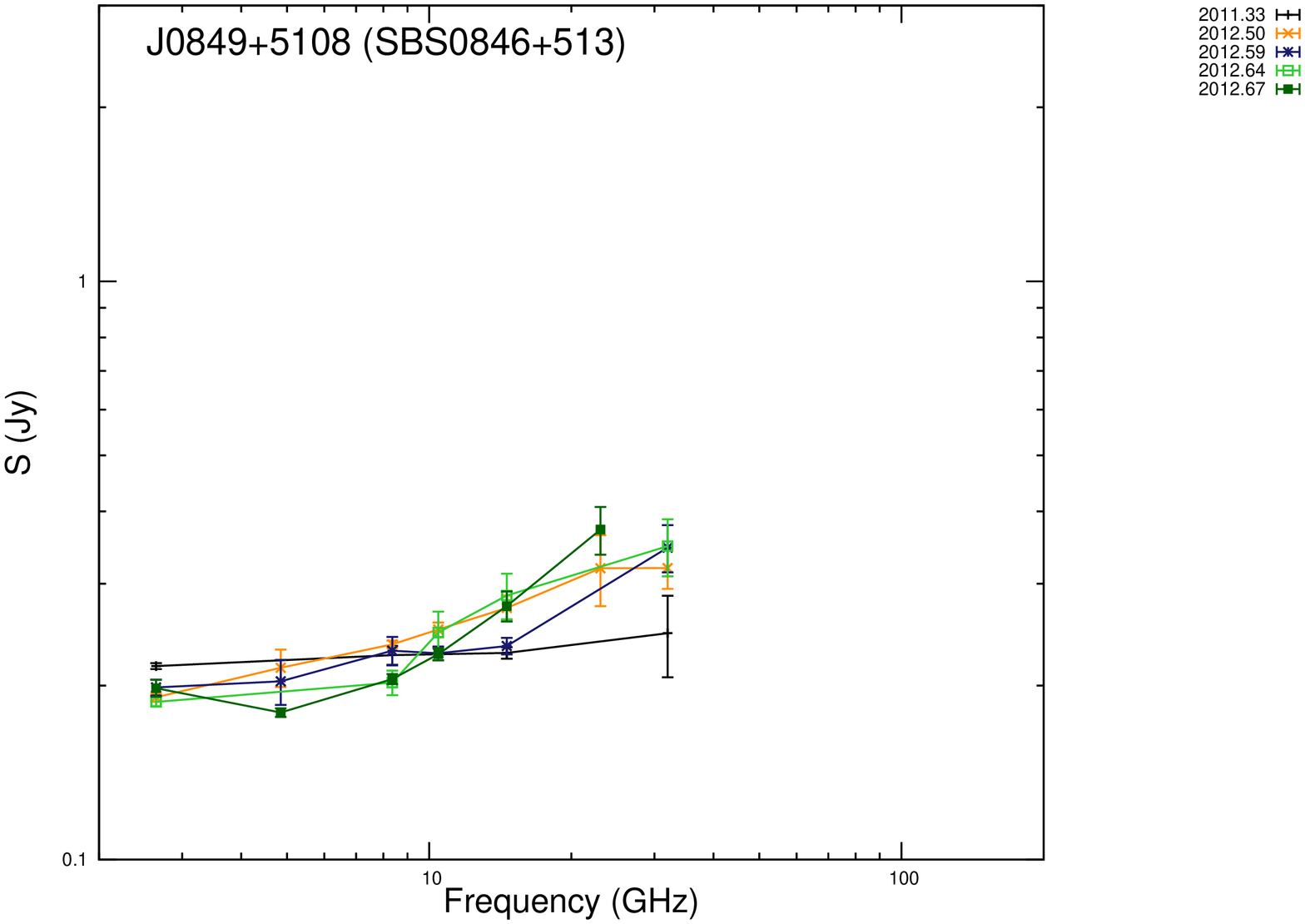}\\
\includegraphics[width=5cm,angle=-90]{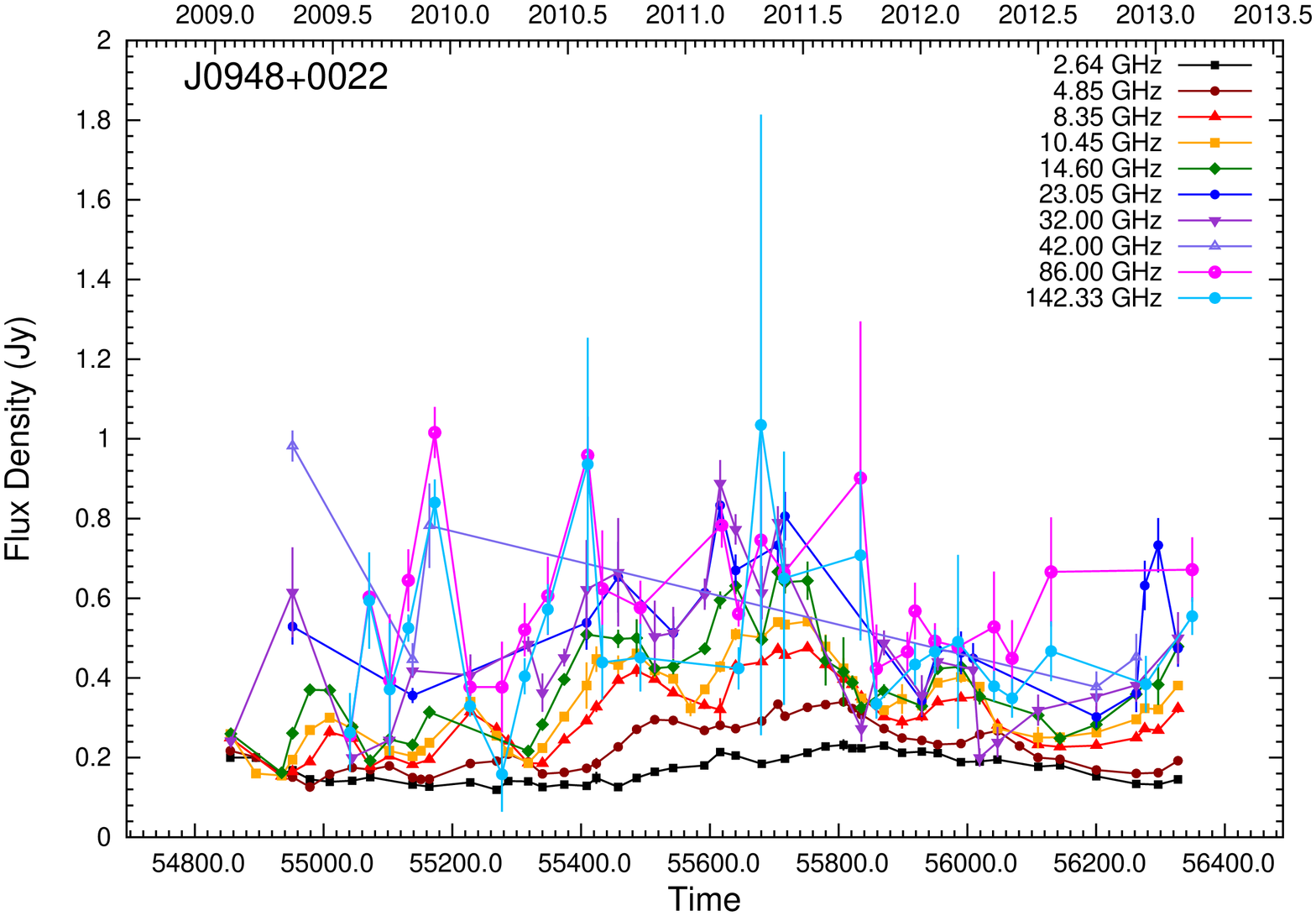}&
\includegraphics[width=5cm,angle=-90]{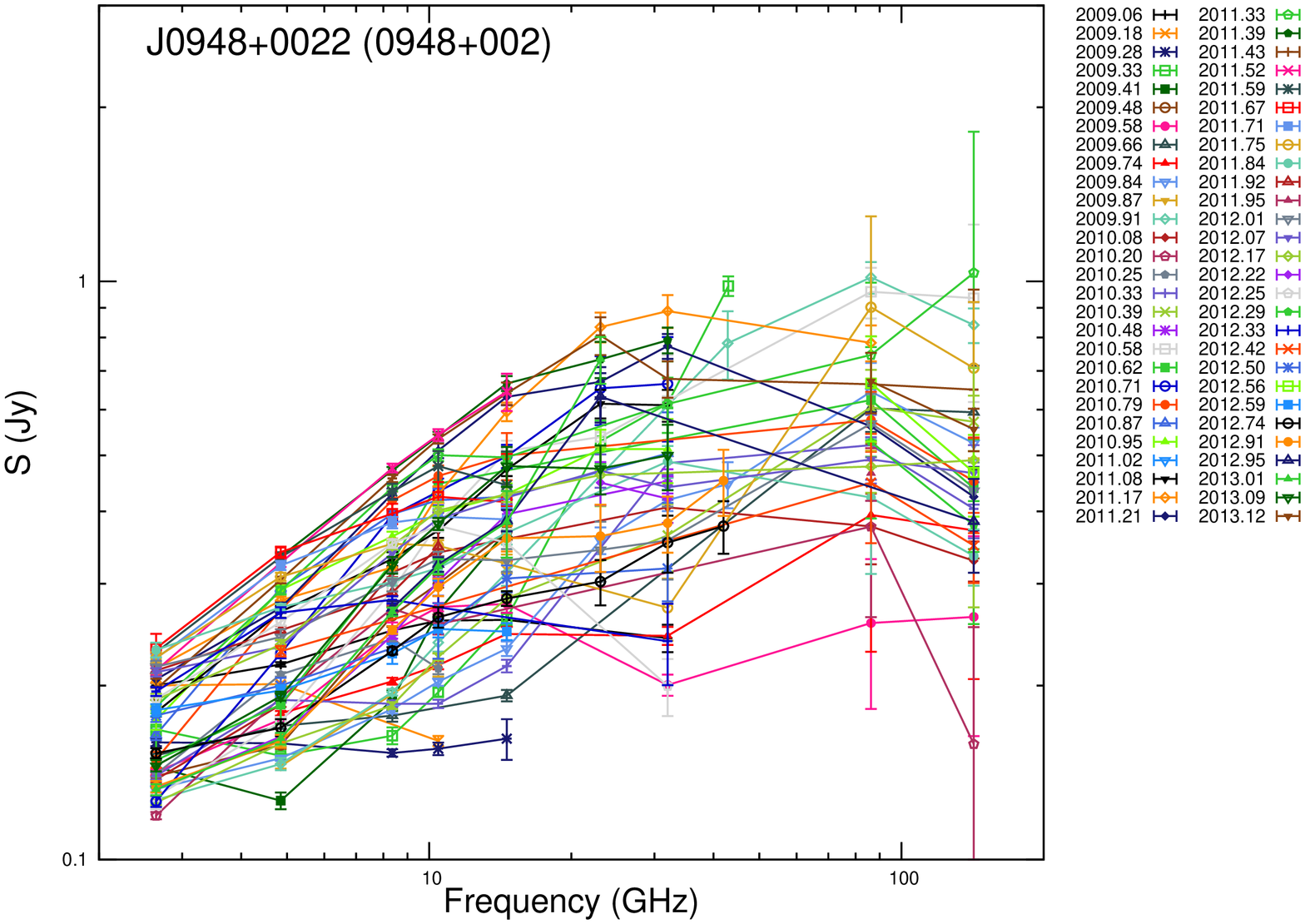}\\
\includegraphics[width=5cm,angle=-90]{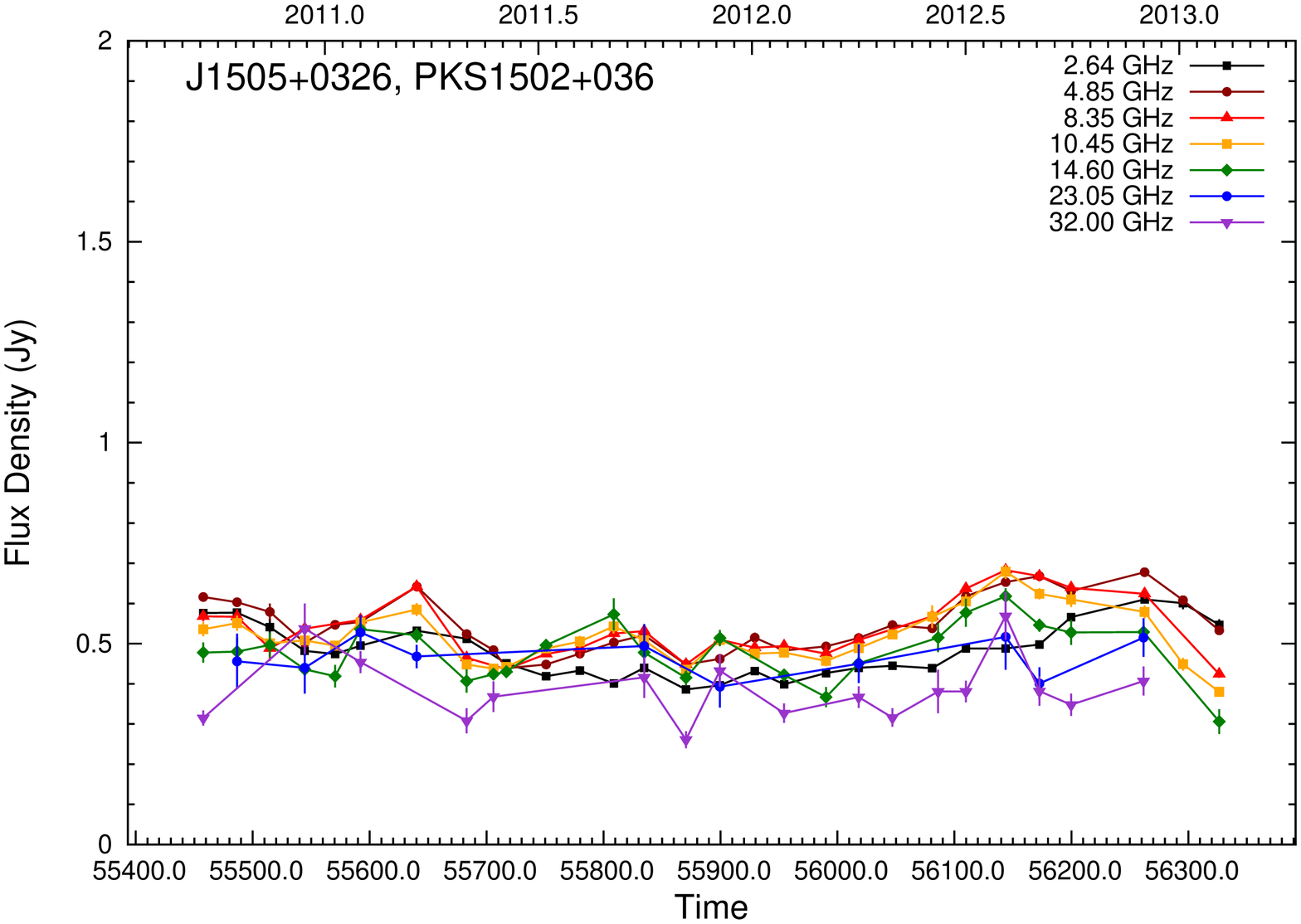}&
\includegraphics[width=5cm,angle=-90]{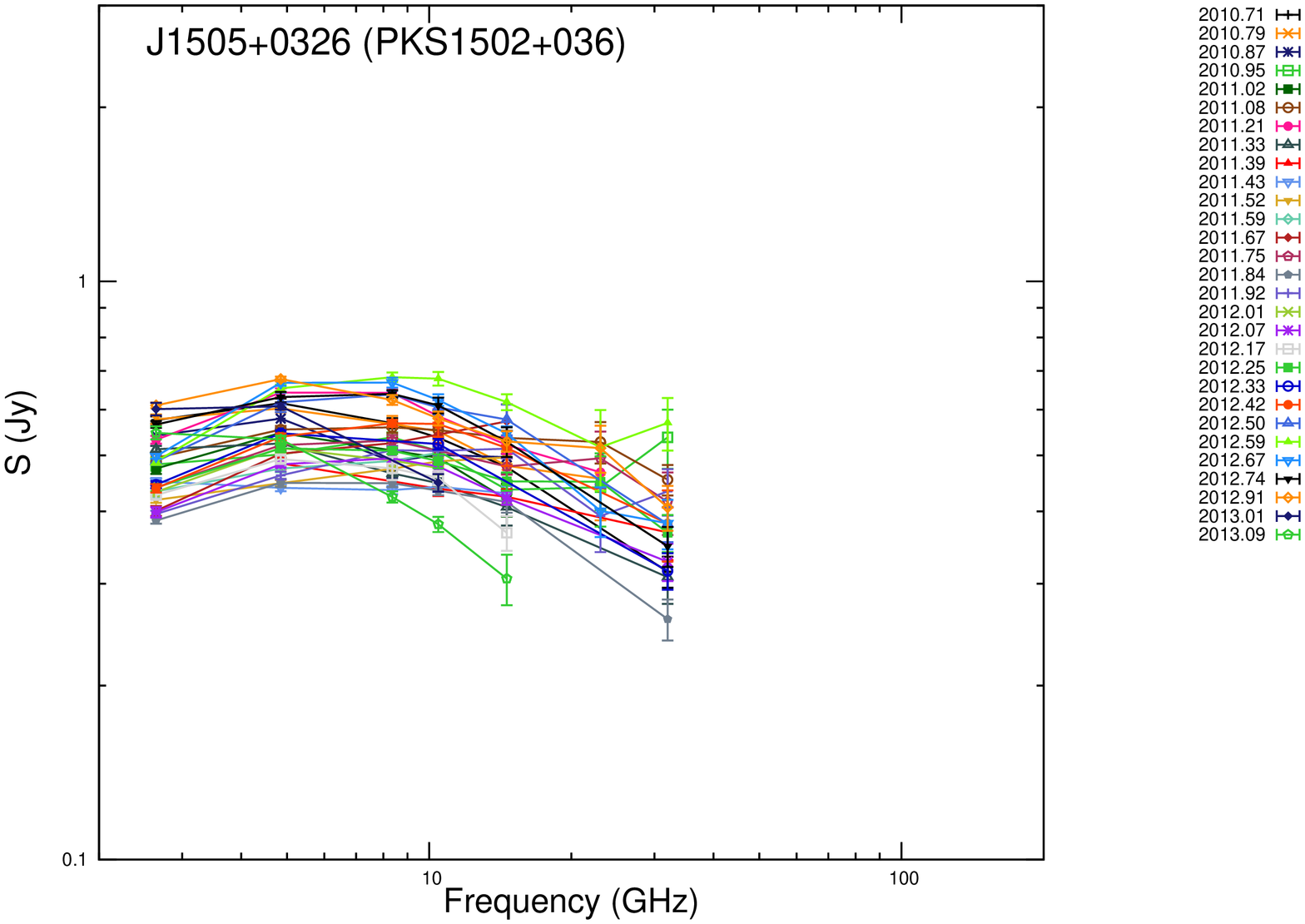}\\
\end{tabular}
\caption{Multi-frequency light curves and radio SEDs of the four monitored NLSy1
  galaxies. Each row corresponds to one source identified by its name in the top-left of
  the corresponding panel. The data points are connected with straight segments only for
  assistance to the eye. The axes limits are kept identical to allow inter-source
  comparisons.}
\label{fig:lc_n_specs}
\end{figure}

Figure~\ref{fig:lc_n_specs} shows the light curves along with the radio SEDs of the four
NLSy1s that have been monitored with Effelsberg and IRAM telescopes. The monitoring program
covers 2.6, 4.85, 8.35, 10.45, 14.60, 23.05, 32 and 43\,GHz with the Effelsberg 100-m
telescope. For the two cases of J0324$+$3410 and J0948$+$0022 detectable with the IRAM
30-m telescope, also data at 86 and 142\,GHz are available. The mean cadence of the light
curves and spectra shown there is approximately 1.5 months. Table~\ref{tab:summary}
reports some indicative light curve parameters.

\begin{table}[h!]
  \caption{Mean flux density and corresponding StDev for each observing
    frequency and source.}     
  \label{tab:summary}  
\footnotesize
  \centering                    
  \begin{tabular}{llcccccccccc} 
    \hline\hline                 
Source         &$\nu^{1}$ &2.64       &4.85  &8.35 &10.45 &14.6 &23.05 &32  &43 &86 &142   \\
\hline            
J0324$+$3410                &<S>$^{2}$ &0.438  &0.373  &0.354  &0.354  &0.376 &0.579  &0.458  &0.302  &0.496 & 0.446\\      
                         &StDev$^{3}$ &0.048  &0.048  &0.057  &0.067  &0.124  &0.250
                         &0.360  &-  &0.170 &0.161 \\      
J0849$+$5108                           &<S> &0.198  &0.199  &0.220  &0.238  &0.259
                         &0.346  &0.315  &-  &- &- \\      
                         &StDev &0.011  &0.018  &0.015  &0.012  &0.026  &0.037
                         &0.048  &-  &- &- \\      
 J0948$+$0022                      &<S> &0.171  &0.224  &0.301  &0.336  &0.390
                         &0.551  &0.469  &0.608  &0.590 &0.502 \\      
                         &StDev &0.035  &0.061  &0.089  &0.104  &0.128  &0.159
                         &0.182  &0.262  &0.183 &0.204 \\      
J1505$+$0326                         &<S> &0.482  &0.546  &0.538  &0.516  &0.479
                         &0.466  &0.386  &-  &- &- \\      
                         &StDev &0.066  &0.070  &0.075  &0.067  &0.071  &0.047
                         &0.080  &-  &- &- \\      
    \hline 
\multicolumn{12}{l}{$^{1}$Frequency in GHz.}                 \\                
\multicolumn{12}{l}{$^{2}$Mean flux density in Jy.}                       \\          
\multicolumn{12}{l}{$^{3}$Standard Deviation of around the mean flux density in
  Jy: measure of the variability amplitude. }\\                                 
  \end{tabular}
\end{table}
 Apart from the case of J0849$+$5108 for which the dataset does not cover a 
substantial number of activity cycles, the remaining three sources show intense
variability at practically all frequencies, a behaviour systematically seen in
blazars. The variability amplitude is frequency dependent, following the typical fashion:
higher frequencies vary more, which comprises yet another indication that variability
mechanisms common with those acting in typical blazars, are present. 

The phenomenologies seen in the light curves of these sources vary significantly and
chiefly in terms of the observed power of the different events. In these terms
J1505$+$0326 shows very weak flaring events. J0324$+$3410 shows at least two major events present
at the high-end of the bandpass which disappear at lower frequencies. J0948$+$0022 show
frequent events of activity which happen considerably fast. 

\subsection{Variability brightness temperatures and Doppler factors}
Under the condition of long enough datasets, the previously discussed light curves can
provide some insights into the energetics of the emitting regions responsible for the
outbursts. From the observed variability amplitude and the time over which the
event spans, one can work out 
the brightness temperature associated with the event, $T_\mathrm{B}$. In
table~\ref{tab:tbs} are reported the calculated maximum brightness temperatures at two
characteristic frequencies, namely 8.35 (low band) and 32\,GHz (high band). 
Imposing an equipartition brightness temperature upper limit of $5\cdot10^{10}$\,K
\citep{Readhead1994ApJ}, one can estimate the Doppler factors required to explain the
observed excess.
The levels of $T_\mathrm{B}$ seen in the three studied NLSy1s are rather moderate -- compared
to those usually seen in blazars -- and closer to those of the lower luminosity blazars,
namely the BL Lacs. Consequently, even at the high frequencies the inferred Doppler factors
are moderate.
\begin{table}[h!]
  \caption{Variability brightness temperatures and Doppler factors.}     
  \label{tab:tbs}  
  \footnotesize
  \centering                    
  \begin{tabular}{cccccccccc} 
    \hline\hline                 
   Source &\multicolumn{4}{c}{8.35\,GHz}& &\multicolumn{4}{c}{32\,GHz}  \\
\cline{2-5}
\cline{7-10}                   
         &$\Delta S$ &$\Delta t$ &$T_\mathrm{B, max}$ &$D$ & &$\Delta S$ &$\Delta t$    &$T_\mathrm{B, max}$ &$D$ \\
         &(Jy) &(d) &(K) & & &(Jy) &(d)  &(K) & \\
    \hline         
    J0324$+$3410 &0.06  &180 &$6\cdot10^{10}$  &1 & &0.5   &150.0 &$5\cdot10^{10}$ &1  \\
    J0948$+$0022 &0.07  &150 &$4\cdot10^{12}$  &5 & &0.2   &320.0 &$2\cdot10^{11}$ &2 \\
    J1505$+$0326 &0.08  &180 &$2\cdot10^{12}$  &4 & &0.09 &120.0 &$3\cdot10^{11}$ &2 \\
    \hline 
  \end{tabular}
\end{table}
%
%

\section{Broad-band spectra variability}

Figure~\ref{fig:lc_n_specs} shows the time sampled broad-band spectra. The combined
Effelsberg-IRAM spectra are simultaneous within 1.1 days securing that they are free of source
variability noise. The mean sampling is one spectrum every 1.5 months (see also
Sect.~\ref{sec:lcs}).

Except for J1505$+$0326 where the spectral evolution is not discernible, all other
three NLSy1 would be classified as type ``1'' or ``2'' in the classification proposed by
\cite{2012JPhCS.372a2007A} showing intense spectral evolution. 

Despite the limited dataset, J0849$+$5108 behaves qualitatively similarly to
J0948$+$0022. Both show very hard spectra with practically a complete absence of signs of
a quiescence spectrum. The spectra are highly inverted and intensely variable. In the case
of J0948$+$0022 the source shows an impressive spectral evolution and remains at the
hard-spectrum state (i.e. $\alpha\ge 0.0$ with $S\sim\nu^{\alpha}$) 4\% of the time and
its spectral index can reach values of $+1.0$; an indication of high frequency components
dominating the observed emission. J0849$+$5108, on the other hand, shows flat spectra with
a spectral index not exceeding values of the order of $+0.2$.

In the case of J0324$+$3410 a steep spectrum component emerges at frequencies below
10.45 GHz. It appears intensely variable with the variability events transversing the
bandpass, leaving however a ``naked'' steep spectrum behind. The low-band spectral index
spends less than 3\,\% of the time in negative values and can get as steep as $-0.4$, very
close to the commonly assumed values of $-0.5$ or $-0.7$; yet an indication of a large
scale jet. The significant linear polarisation seen in the source also supports the idea
of a significant contribution from a jet as it is discussed in Sect.~\ref{sec:pol}.

J1505$+$0326 however imitates the behaviour of type ``5b'' in \cite{2012JPhCS.372a2007A}
with only mild spectral evolution.
Its spectrum is mostly convex, and at times shows high frequency components
expanding fast towards lower bands.

\section{Linear polarisation}
\label{sec:pol}
Assuming that the variability seen in all the light curves discussed
here is attributed to synchrotron self-absorbed components
it is expected that significant linear polarisation is detected
during their optically thin phase. The degree of polarisation is a
function of the spectral index. Assuming highly ordered and
intense magnetic fields \citep{1967Natur.216..147S}, as those
expected during flares, even circular polarisation could occur. The
superposition however of different components will result in ``beam
depolarisation''.  

 Table~\ref{tab:pol} summarises the mean fractional linear polarisation
$<\Pi>$ and its standard deviation StDev for the four NLSy1s
discussed here at 8.35\,GHz.
\begin{figure}[] 
\centering
\begin{tabular}{c}
\includegraphics[trim=3.2cm 1cm 9.7cm 1cm,clip=true,totalheight=0.38\textheight,angle=-90]{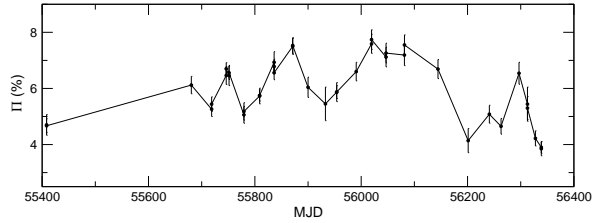} 
\end{tabular}
\caption{The fractional linear polarisation at 8.35\,GHz for
  J0324$+$3410. The effects of instrumental polarisation are, to first
  order, ignored but only significant detections (SNR better than 3)
  of polarised flux have been used. }
\label{fig:0324_pol}
\end{figure}
As it can be seen there, all four sources display detectable polarisation, though of
relatively low values. J0324$+$3410 however shows values larger than those typically
observed in AGN at these frequency bands ($\sim4$\,\% at 5\,GHz,
e.g. \citealt{Klein2003AnA}).  Figure~\ref{fig:0324_pol} shows its fractional linear
polarisation light curve at 8.35\,GHz. Its mean is of the order of 6\,\% while at
4.85\,GHz it reaches values of 6.9\,\%. This phenomenology could be explained by assuming
that the optically thin emission dominates the high frequency component (HFC) towards
lower frequencies. The 4.85-GHz emission is less ``contaminated'' by the optically thick
emission introduced by the HFC. It remains optically thin hence contributing
higher fractional polarisation. Similar arguments hold for J0849$+$5108 although the
effect is much more moderate.
\begin{table}[h!]
  \caption{Mean fractional linear polarisation at 8.35\,GHz of the
    four studied NLSy1s.}     
  \label{tab:pol}  
\footnotesize
  \centering                    
  \begin{tabular}{llcccc} 
    \hline\hline                 
Source    &      &J0324$+$3410 &J0849$+$5108  &J0948$+$0022 &J1505$+$0326  \\
\hline         
$<\Pi>$ &(\%)             &6.0 &2.7 &1.5 &1.3\\       
StDev &(\%)             &1.1 &0.6 &0.5&0.5 \\
    \hline 
  \end{tabular}
\end{table}

Unlike J0324$+$3410, the other NLSy1s show very low polarisation
degree. Their spectral variability pattern indicates that the sources
undergo flaring events which ``show'' their optically thick part at
the investigated frequency of 8.35\,GHz which is of low ``emitted''
polarisation. Polarisation measurements above 43\,GHz and above
10.45\,GHz for J0948$+$0022 and J1505$+$0326, respectively, should
disclose a significant degree of polarisation.

\section{Discussion and conclusions}
This contribution  is meant as a summary of the first results of the radio monitoring of four
$\gamma$-loud NLSy1s. As expected, the sources show typical blazar-like behaviour with
variable emission at practically all frequencies as already discussed by
\cite{2011nlsg.confE..26F}. Outbursts causing changes in the flux density with
factors up to 5 are seen. The variability brightness temperatures however do not
dramatically exceed the theoretically imposed upper limits (equipartition). They imply
rather moderate equipartition Doppler factors possibly associated with the systematically
smaller black hole masses seen in these systems \citep{2006AJ....132..531K}. In the
frequency domain on the other hand, the sources show very intense variability. Except for
J1505$+$0326 which displays moderate evolution, the rest would be classified as type 1 or 2
according to \citet{2012JPhCS.372a2007A}. The variability pattern indicates variability
mechanisms similar to those acting in the jets of blazars. Specifically for J0948$+$0022
the evolution happens remarkably fast making it an excellent laboratory for the study of
the spectral evolution of the high frequency component. The emergence of a steep spectrum
component towards low bands for J0324$+$3410 and J0849$+$5108 and the increase of linear
polarisation is yet another indication for the presence of an optically thin quiescent jet
similar to that seen in typical blazars. A careful cross-correlation of the spectral index
light curves with the fractional polarisation is underway to investigate the exact
connection between the two and compare it with the synchrotron theory.

\begin{multicols}{2}
\footnotesize
\bibliographystyle{aa} 
\bibliography{/Users/mangel/work/Literature/MyBIB/References.bib} 
\end{multicols}



\end{document}